\documentstyle[aps,prl,multicol,epsf,graphics]{revtex}

\title{A Linear Optics Implementation of Weak Values in Hardy's Paradox}

\author{S. E. Ahnert and M. C. Payne}
\address{Theory of Condensed Matter Group, Cavendish Laboratory, \\
Madingley Road, Cambridge CB3 0HE, U.K.}

\begin{document}
\maketitle 
\begin{abstract}

We propose an experimental setup for the implementation of weak measurements in the context of the gedankenexperiment known as Hardy's Paradox. As Aharonov {\em et al.} showed, these weak values form a language with which the paradox can be resolved. Our analysis shows that this language is indeed consistent and experimentally testable. It also reveals exactly how a combination of weak values can give rise to an apparently paradoxical result.

\bigskip

\end{abstract}

\begin{multicols}{2}

\section{introduction}
Hardy's paradox, introduced in 1992 \cite{hardy}, is a strong proof of non-locality in quantum mechanics. It has been generalized as a test of non-locality by Hardy in a later paper \cite{hardy93} and has subsequently been tested using photons \cite{mandel,demartini}. A decade after the original paper, Aharonov {\em et al.} \cite{popescu} published an analysis of the original paradox from the perspective of weak values. Weak measurements, in which such values arise, are a generalization of the quantum mechanical measurement process including uncertain pointer positions, and were first introduced by Aharonov, Albert and Vaidman (AAV) in 1988\cite{aharonov88}. Using these weak values, the authors of \cite{popescu} are able to recast the counterfactual statements of the paradox in a more consistent language. The aim of our paper is to suggest a feasible implementation of these weak measurements. In doing so we also reveal more clearly how these weak values give rise to a seemingly paradoxical outcome.

\section{Hardy's Paradox}

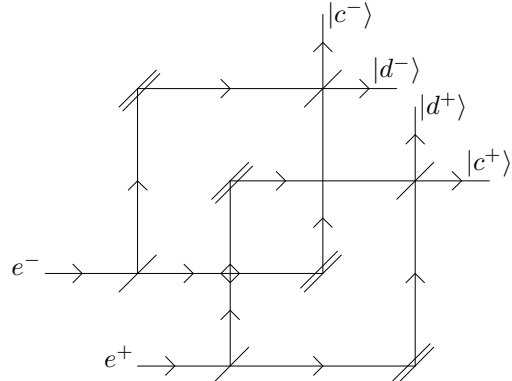
\begin{figure}
\setlength{\unitlength}{0.7pt}
\ifx\plotpoint\undefined\newsavebox{\plotpoint}\fi
\begin{picture}(250,200)(0,0)
\font\gnuplot=cmr10 at 10pt
\gnuplot
\put(50,50){\special{em:moveto}}
\put(70,70){\special{em:lineto}}
\put(150,150){\special{em:moveto}}
\put(170,170){\special{em:lineto}}
\put(150,50){\special{em:moveto}}
\put(170,70){\special{em:lineto}}
\put(148,52){\special{em:moveto}}
\put(168,72){\special{em:lineto}}
\put(50,150){\special{em:moveto}}
\put(70,170){\special{em:lineto}}
\put(52,148){\special{em:moveto}}
\put(72,168){\special{em:lineto}}
\put(10,60){\special{em:moveto}}
\put(160,60){\special{em:lineto}}
\put(60,60){\special{em:moveto}}
\put(60,160){\special{em:lineto}}
\put(60,160){\special{em:moveto}}
\put(200,160){\special{em:lineto}}
\put(160,60){\special{em:moveto}}
\put(160,200){\special{em:lineto}}
\put(25,55){\special{em:moveto}}
\put(30,60){\special{em:lineto}}
\put(25,65){\special{em:lineto}}
\put(85,55){\special{em:moveto}}
\put(90,60){\special{em:lineto}}
\put(85,65){\special{em:lineto}}
\put(155,85){\special{em:moveto}}
\put(160,90){\special{em:lineto}}
\put(165,85){\special{em:lineto}}
\put(105,155){\special{em:moveto}}
\put(110,160){\special{em:lineto}}
\put(105,165){\special{em:lineto}}
\put(55,105){\special{em:moveto}}
\put(60,110){\special{em:lineto}}
\put(65,105){\special{em:lineto}}
\put(155,180){\special{em:moveto}}
\put(160,185){\special{em:lineto}}
\put(165,180){\special{em:lineto}}
\put(180,155){\special{em:moveto}}
\put(185,160){\special{em:lineto}}
\put(180,165){\special{em:lineto}}

\put(100,0){\special{em:moveto}}
\put(120,20){\special{em:lineto}}
\put(200,100){\special{em:moveto}}
\put(220,120){\special{em:lineto}}
\put(200,0){\special{em:moveto}}
\put(220,20){\special{em:lineto}}
\put(198,2){\special{em:moveto}}
\put(218,22){\special{em:lineto}}
\put(100,100){\special{em:moveto}}
\put(120,120){\special{em:lineto}}
\put(102,98){\special{em:moveto}}
\put(122,118){\special{em:lineto}}
\put(60,10){\special{em:moveto}}
\put(210,10){\special{em:lineto}}
\put(110,10){\special{em:moveto}}
\put(110,110){\special{em:lineto}}
\put(110,110){\special{em:moveto}}
\put(250,110){\special{em:lineto}}
\put(210,10){\special{em:moveto}}
\put(210,150){\special{em:lineto}}
\put(75,5){\special{em:moveto}}
\put(80,10){\special{em:lineto}}
\put(75,15){\special{em:lineto}}
\put(155,5){\special{em:moveto}}
\put(160,10){\special{em:lineto}}
\put(155,15){\special{em:lineto}}
\put(205,55){\special{em:moveto}}
\put(210,60){\special{em:lineto}}
\put(215,55){\special{em:lineto}}
\put(135,105){\special{em:moveto}}
\put(140,110){\special{em:lineto}}
\put(135,115){\special{em:lineto}}
\put(105,35){\special{em:moveto}}
\put(110,40){\special{em:lineto}}
\put(115,35){\special{em:lineto}}
\put(205,130){\special{em:moveto}}
\put(210,135){\special{em:lineto}}
\put(215,130){\special{em:lineto}}
\put(230,115){\special{em:moveto}}
\put(235,110){\special{em:lineto}}
\put(230,105){\special{em:lineto}}
\put(110,65){\special{em:moveto}}
\put(115,60){\special{em:lineto}}
\put(110,55){\special{em:lineto}}
\put(105,60){\special{em:lineto}}
\put(110,65){\special{em:lineto}}
\put(0,65){\makebox(0,0){$e^-$}}
\put(50,15){\makebox(0,0){$e^+$}}
\put(175,200){\makebox(0,0){$| c^- \rangle$}}
\put(200,170){\makebox(0,0){$| d^- \rangle$}}
\put(225,150){\makebox(0,0){$| d^+ \rangle$}}
\put(250,120){\makebox(0,0){$| c^+ \rangle$}}
\end{picture}
\caption{The original setup of Hardy's paradox. It comprises two interleaving Mach-Zehnder interferometers, traversed simultaneously by a positron and an electron respectively, and arranged in such a way that two arms overlap. As a result the electron and positron cannot both travel through the overlapping arm of the respective interferometer without annihilating. The removal of any one or both the final beamsplitters in this setup leads to paradoxical results which serve as a proof of quantum mechanical non-locality.}\label{hardyfig} 
\end{figure}

In Hardy's original setup, an electron and a positron each traverse a Mach-Zehnder interferometer. The two interferometers interleave in such a way that with a probability of ${1 \over 4}$ the two particles can be found at the same position and thus annihilate. The setup can be seen in Fig. \ref{hardyfig}. 
Hardy then proceeds to calculate the combined states of the particles after they have left the interferometers for four different cases: 

\begin{enumerate}
\item{The second (reunifying) beamsplitters are present in both interferometers, i.e. both interferometers are complete.}
\item{The positron's second beamsplitter is absent, the electron's is present.}
\item{The electron's second beamsplitter is absent, the positron's is present.}
\item{Both second beamsplitters are absent.}
\end{enumerate}

The final two-particle states for the four cases are:

\begin{eqnarray}\label{heq1}
| \Psi_1 \rangle = {1 \over 4} [-2 | \gamma \rangle - 3 | c^+ \rangle | c^- \rangle + i | c^+ \rangle | d^- \rangle 
\cr
+ i | d^+ \rangle | c^- \rangle - | d^+ \rangle | d^- \rangle] 
\end{eqnarray}
\begin{equation}\label{heq2}
| \Psi_2 \rangle = {1 \over 2 \sqrt{2}} [ - \sqrt{2} | \gamma \rangle - | c^+ \rangle | c^- \rangle + i | c^+ \rangle | d^- \rangle + 2 i | d^+ \rangle | c^- \rangle] 
\end{equation}
\begin{equation}\label{heq3}
| \Psi_3 \rangle = {1 \over 2 \sqrt{2}} [ - \sqrt{2} | \gamma \rangle - | c^+ \rangle | c^- \rangle + 2 i | c^+ \rangle | d^- \rangle + i | d^+ \rangle | c^- \rangle] 
\end{equation}
\begin{equation}\label{heq4}
| \Psi_4 \rangle = {1 \over 2} [- | \gamma \rangle + i | c^+ \rangle | d^- \rangle + i | d^+ \rangle | c^- \rangle + | d^+ \rangle | d^- \rangle]
\end{equation}

where $| c^\pm \rangle$ and $| d^\pm \rangle$ denote the different exit path states respectively, of the positron (+) and electron (-) from the interferometers, and $\gamma$ denotes the annihilation event. 

If locality is assumed, then, as Hardy argues, the result of a measurement on one particle should not depend on the choice of measurement for another particle. Thus one can construct the quantities $C^\pm(0)$ and $D^\pm(\infty)$ which both take values of either one or zero and thus denote the events of measuring a particle (= 1) or not (= 0) for exit path states $| c \rangle$ and $| d \rangle$ of the electron and positron. The values of 0 and $\infty$ indicate whether the beamsplitters are present (0) or have been removed ($\infty$). Thus the following argument can be made:

From equation \ref{heq4} it follows that:

\begin{equation}
C^+(\infty)C^-(\infty) = 0
\end{equation}

Whereas equation \ref{heq3} and \ref{heq2} imply:

\begin{equation}
D^+(0) = 1 \Rightarrow C^-(\infty) = 1
\end{equation}

\begin{equation}
D^-(0) = 1 \Rightarrow C^+(\infty) = 1
\end{equation}

respectively. But according to equation (\ref{heq1}), in ${1 \over 16}$th of experiments the outcomes would indicate that:

\begin{equation}
D^+(0)D^-(0) = 1
\end{equation}

Hence, in these experiments there exists a contradiction, as there are no local quantities $C^\pm(\infty)$ and $D^\pm(0)$ which can fulfill all these equations.
\section{Weak measurement interpretation of Hardy's Paradox}

In a recent paper, Aharonov {\em et al.} \cite{popescu} offer an interpretation of Hardy's paradox using weak values. Weak measurement was introduced in 1988 by Aharonov, Albert and Vaidman (AAV) and is a generalization of quantum mechanics to uncertain measurement pointers. Such pointers, combined with post-selection of a specific quantum state can give measurement results far outside the spectrum of eigenvalues. These values however are not just mere statistical errors, but are consistent and have a physical meaning. An often-quoted example is the weak localization of a particle inside a barrier (where it would never be found in a 'strong' measurement) which yields the correct value of negative kinetic energy \cite{negkin}. 
Weak values are calculated using the expression

\begin{eqnarray}
A_w = {\langle \Psi_f | A | \Psi_i \rangle \over \langle \Psi_f | \Psi_i \rangle} 
\end{eqnarray}

where A is the operator of the observable to be measured weakly, $| \Psi_i \rangle$ is the initial state of the system and $| \Psi_f \rangle$ is the post-selected state.

In \cite{popescu} the authors calculate weak particle occupation numbers of the different branches in Hardy's interferometric setup. Their initial state is combined state of the electron and positron after the annihilation point, i.e.:

\begin{equation}
| \Psi_i \rangle = {1 \over \sqrt{3}} [| O^+ \rangle | NO^- \rangle + | NO^+ \rangle | O^- \rangle + | NO^+ \rangle | NO^- \rangle]
\end{equation}

and their post-selected state corresponds to the simultaneous firing of detectors $D^+$ and $D^-$:

\begin{equation}
| \Psi_f \rangle = {1 \over \sqrt{2}} [| NO^+ \rangle - | O^+ \rangle] [| NO^- \rangle - | O^- \rangle]
\end{equation}

where $| O^\pm \rangle$ and $| NO^\pm \rangle$ correspond to path states in the overlapping and non-overlapping arms respectively, of the positron (+) and electron (-) interferometers. 

The weak values for single-particle occupation numbers turn out to be:

\begin{eqnarray}\label{pon}
N^-_{Ow} = 1 \cr
N^+_{Ow} = 1 \cr
N^-_{NOw} = 0 \cr
N^+_{NOw} = 0 \cr
\end{eqnarray}

and the two-particle occupation numbers are:

\begin{eqnarray}\label{pon2}
N^{+,-}_{O,Ow} = 0 \cr
N^{+,-}_{O,NOw} = 1 \cr
N^{+,-}_{NO,Ow} = 1 \cr
N^{+,-}_{NO,NOw} = -1 \cr
\end{eqnarray}

where the last value is the most significant, as it points to the possibility of having weak negative particle occupation numbers. This is discussed in some detail in \cite{popescu}. 

The aim of our work is propose an implementation of these weak measurements in order to clarify how these values arise, what their physical meaning is and how they interact to give a seemingly paradoxical outcome. 

\section{Weak values of arrival time}

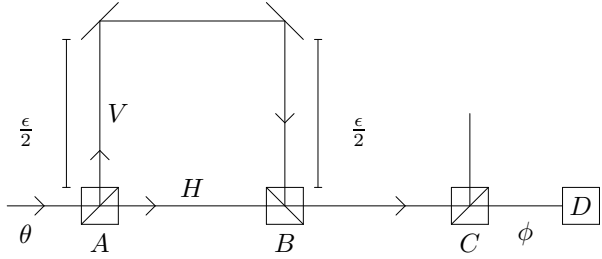
\begin{figure}

\setlength{\unitlength}{0.7pt}
\ifx\plotpoint\undefined\newsavebox{\plotpoint}\fi
\begin{picture}(300,180)(0,0)
\font\gnuplot=cmr10 at 10pt
\gnuplot
\put(50,50){\special{em:moveto}}
\put(70,50){\special{em:lineto}}
\put(70,70){\special{em:lineto}}
\put(50,70){\special{em:lineto}}
\put(50,50){\special{em:lineto}}
\put(70,70){\special{em:lineto}}
\put(330,50){\special{em:moveto}}
\put(310,50){\special{em:lineto}}
\put(310,70){\special{em:lineto}}
\put(330,70){\special{em:lineto}}
\put(330,50){\special{em:lineto}}
\put(250,50){\special{em:moveto}}
\put(270,50){\special{em:lineto}}
\put(270,70){\special{em:lineto}}
\put(250,70){\special{em:lineto}}
\put(250,50){\special{em:lineto}}
\put(270,70){\special{em:lineto}}
\put(150,70){\special{em:moveto}}
\put(170,70){\special{em:lineto}}
\put(170,50){\special{em:lineto}}
\put(150,50){\special{em:lineto}}
\put(150,70){\special{em:lineto}}
\put(170,50){\special{em:lineto}}
\put(150,170){\special{em:moveto}}
\put(170,150){\special{em:lineto}}
\put(50,150){\special{em:moveto}}
\put(70,170){\special{em:lineto}}
\put(10,60){\special{em:moveto}}
\put(310,60){\special{em:lineto}}
\put(260,60){\special{em:moveto}}
\put(260,110){\special{em:lineto}}
\put(60,60){\special{em:moveto}}
\put(60,160){\special{em:lineto}}
\put(60,160){\special{em:moveto}}
\put(160,160){\special{em:lineto}}
\put(160,60){\special{em:moveto}}
\put(160,160){\special{em:lineto}}
\put(25,55){\special{em:moveto}}
\put(30,60){\special{em:lineto}}
\put(25,65){\special{em:lineto}}
\put(85,55){\special{em:moveto}}
\put(90,60){\special{em:lineto}}
\put(85,65){\special{em:lineto}}
\put(155,110){\special{em:moveto}}
\put(160,105){\special{em:lineto}}
\put(165,110){\special{em:lineto}}
\put(55,85){\special{em:moveto}}
\put(60,90){\special{em:lineto}}
\put(65,85){\special{em:lineto}}
\put(220,55){\special{em:moveto}}
\put(225,60){\special{em:lineto}}
\put(220,65){\special{em:lineto}}
\put(40,70){\special{em:moveto}}
\put(44,70){\special{em:lineto}}
\put(42,70){\special{em:moveto}}
\put(42,150){\special{em:lineto}}
\put(44,150){\special{em:moveto}}
\put(40,150){\special{em:lineto}}
\put(180,70){\special{em:moveto}}
\put(176,70){\special{em:lineto}}
\put(178,70){\special{em:moveto}}
\put(178,150){\special{em:lineto}}
\put(176,150){\special{em:moveto}}
\put(180,150){\special{em:lineto}}
\put(320,60){\makebox(0,0){$D$}}
\put(20,45){\makebox(0,0){$\theta$}}
\put(60,40){\makebox(0,0){$A$}}
\put(160,40){\makebox(0,0){$B$}}
\put(110,70){\makebox(0,0){$H$}}
\put(70,110){\makebox(0,0){$V$}}
\put(290,45){\makebox(0,0){$\phi$}}
\put(260,40){\makebox(0,0){$C$}}
\put(200,100){\makebox(0,0){${\epsilon \over 2}$}}
\put(20,100){\makebox(0,0){${\epsilon \over 2}$}}
\end{picture}

\caption{The photon, incident with polarization angle $\theta$, is split into its horizontal (H) and vertical (V) polarization components at a polarizing beamsplitter (A). These then reach the second polarizing beamsplitter (B) at different times with $\Delta t = {\epsilon \over c}$. Thus the arrival time has become entangled with the polarization which together with the post-selection achieved by the third polarizing beamsplitter (C), corresponds to a weak measurement. However, this is only the case if the uncertainty in arrival time is similar in size to $\epsilon$, the separation of the wavepackets. In other words, $\epsilon \simeq {c \over 4 \pi \Delta \nu}$, where $\Delta \nu$ is the uncertainty in frequency. The basis of beamsplitter C is rotated by an angle $\phi$.}\label{wmfig}
\end{figure}

In earlier work \cite{us} we introduced a setup for implementing weak values of the arrival times of single photons and pairs of entangled photons. In this setup, (Fig. \ref{wmfig}) polarization and arrival time are entangled, after which a particular polarization state is post-selected. Thus weak values of arrival time arise, which due to the entanglement between arrival time and polarization can be interpreted as pointer positions of a measurement device measuring polarization weakly. This effect is very closely related to the explanation of anomalous pulse delay given by Gisin {\em et al.} \cite{gisin}.

\section{Hardy's paradox with weak measurements}

In order to implement Hardy's original paradox \cite{hardy} with photons we need to entangle two photons non-maximally in a state of the form:

\begin{eqnarray}
| \Psi \rangle = {1 \over \sqrt{3}} [| H H \rangle + | H V \rangle + | V H \rangle]
\end{eqnarray}

This we achieve by using a variant of entanglement swapping \cite{zeilinger}, which entangles two non-interacting photons. Initially, two pairs of photons are created together, each pair being in the maximally entangled state

\[
{1 \over \sqrt{2}} \left(| HH \rangle + | VV \rangle\right)
\]

One member of each pair traverses a polarizing beamsplitter (which reflects the vertical polarization component and transmits the horizontal) before its path is combined - in a conventional beamsplitter - with the path of its counterpart from the other pair. A detector is placed behind one of the exits of this beamsplitter. If this detector fires, the initial state of the four photons

\begin{eqnarray}\label{initial}
| \Phi \rangle = {1 \over 2} [| H_1 H_2 \rangle | H_3 H_4 \rangle + | V_1 V_2 \rangle | H_3 H_4 \rangle \cr
+ | H_1 H_2 \rangle | V_3 V_4 \rangle + | V_1 V_2 \rangle | V_3 V_4 \rangle]
\end{eqnarray} 

is projected into:

\begin{eqnarray}\label{hardystate}
| \Psi_i \rangle = {1 \over \sqrt{3}} [| H_2 H_4 \rangle + | H_2 V_4 \rangle + | V_2 H_4 \rangle]
\end{eqnarray}

Note that for this to be successful the detector has to be unable to distinguish between the arrival of one and two photons. This is because the first term in (\ref{hardystate}) corresponds to the arrival of two photons, and the other two terms correspond to only one photon being detected. In addition any photons at the other output port of the conventional beamsplitter must not be detected in any case. 
  
\begin{figure}
\setlength{\unitlength}{0.35pt}
\ifx\plotpoint\undefined\newsavebox{\plotpoint}\fi
\begin{picture}(500,380)(-250,0)
\font\gnuplot=cmr10 at 10pt
\gnuplot
\put(50,50){\special{em:moveto}}
\put(70,50){\special{em:lineto}}
\put(70,70){\special{em:lineto}}
\put(50,70){\special{em:lineto}}
\put(50,50){\special{em:lineto}}
\put(70,70){\special{em:lineto}}
\put(330,50){\special{em:moveto}}
\put(310,50){\special{em:lineto}}
\put(310,70){\special{em:lineto}}
\put(330,70){\special{em:lineto}}
\put(330,50){\special{em:lineto}}
\put(250,50){\special{em:moveto}}
\put(270,50){\special{em:lineto}}
\put(270,70){\special{em:lineto}}
\put(250,70){\special{em:lineto}}
\put(250,50){\special{em:lineto}}
\put(270,70){\special{em:lineto}}
\put(150,70){\special{em:moveto}}
\put(170,70){\special{em:lineto}}
\put(170,50){\special{em:lineto}}
\put(150,50){\special{em:lineto}}
\put(150,70){\special{em:lineto}}
\put(170,50){\special{em:lineto}}
\put(150,170){\special{em:moveto}}
\put(170,150){\special{em:lineto}}
\put(50,150){\special{em:moveto}}
\put(70,170){\special{em:lineto}}
\put(-10,60){\special{em:moveto}}
\put(-100,60){\special{em:lineto}}
\put(10,60){\special{em:moveto}}
\put(310,60){\special{em:lineto}}
\put(260,60){\special{em:moveto}}
\put(260,110){\special{em:lineto}}
\put(60,60){\special{em:moveto}}
\put(60,160){\special{em:lineto}}
\put(60,160){\special{em:moveto}}
\put(160,160){\special{em:lineto}}
\put(160,60){\special{em:moveto}}
\put(160,160){\special{em:lineto}}
\put(25,55){\special{em:moveto}}
\put(30,60){\special{em:lineto}}
\put(25,65){\special{em:lineto}}
\put(-25,55){\special{em:moveto}}
\put(-30,60){\special{em:lineto}}
\put(-25,65){\special{em:lineto}}
\put(85,55){\special{em:moveto}}
\put(90,60){\special{em:lineto}}
\put(85,65){\special{em:lineto}}
\put(155,110){\special{em:moveto}}
\put(160,105){\special{em:lineto}}
\put(165,110){\special{em:lineto}}
\put(55,85){\special{em:moveto}}
\put(60,90){\special{em:lineto}}
\put(65,85){\special{em:lineto}}
\put(-55,85){\special{em:moveto}}
\put(-60,90){\special{em:lineto}}
\put(-65,85){\special{em:lineto}}
\put(220,55){\special{em:moveto}}
\put(225,60){\special{em:lineto}}
\put(220,65){\special{em:lineto}}
\put(40,70){\special{em:moveto}}
\put(44,70){\special{em:lineto}}
\put(42,70){\special{em:moveto}}
\put(42,150){\special{em:lineto}}
\put(44,150){\special{em:moveto}}
\put(40,150){\special{em:lineto}}
\put(180,70){\special{em:moveto}}
\put(176,70){\special{em:lineto}}
\put(178,70){\special{em:moveto}}
\put(178,150){\special{em:lineto}}
\put(176,150){\special{em:moveto}}
\put(180,150){\special{em:lineto}}
\put(320,60){\makebox(0,0){\tiny$D$}}
\put(110,70){\makebox(0,0){\tiny$H$}}
\put(70,110){\makebox(0,0){\tiny$V$}}
\put(290,35){\makebox(0,0){\tiny$\phi = - {\pi \over 4}$}}
\put(200,100){\makebox(0,0){\tiny${\epsilon - \gamma \over 2}$}}
\put(20,100){\makebox(0,0){\tiny${\epsilon - \gamma \over 2}$}}
\put(70,38){\special{em:moveto}}
\put(70,42){\special{em:lineto}}
\put(70,40){\special{em:moveto}}
\put(150,40){\special{em:lineto}}
\put(150,38){\special{em:moveto}}
\put(150,42){\special{em:lineto}}
\put(110,30){\makebox(0,0){\tiny${\gamma}$}}
\put(0,60){\makebox(0,0){\tiny$S$}}
\put(10,50){\special{em:moveto}}
\put(-10,50){\special{em:lineto}}
\put(-10,70){\special{em:lineto}}
\put(10,70){\special{em:lineto}}
\put(10,50){\special{em:lineto}}
\put(-50,50){\special{em:moveto}}
\put(-70,50){\special{em:lineto}}
\put(-70,70){\special{em:lineto}}
\put(-50,70){\special{em:lineto}}
\put(-50,50){\special{em:lineto}}
\put(-70,70){\special{em:lineto}}
\put(-90,265){\special{em:moveto}}
\put(-110,255){\special{em:lineto}}
\put(-91,267){\special{em:moveto}}
\put(-111,257){\special{em:lineto}}
\put(-210,160){\special{em:moveto}}
\put(-190,160){\special{em:lineto}}
\put(-60,60){\special{em:moveto}}
\put(-60,110){\special{em:lineto}}
\put(-150,106){\special{em:moveto}}
\put(-150,110){\special{em:lineto}}
\put(-146,110){\special{em:lineto}}
\put(-100,60){\special{em:moveto}}
\put(-220,180){\special{em:lineto}}
\put(50,250){\special{em:moveto}}
\put(70,250){\special{em:lineto}}
\put(70,270){\special{em:lineto}}
\put(50,270){\special{em:lineto}}
\put(50,250){\special{em:lineto}}
\put(70,270){\special{em:lineto}}
\put(330,250){\special{em:moveto}}
\put(310,250){\special{em:lineto}}
\put(310,270){\special{em:lineto}}
\put(330,270){\special{em:lineto}}
\put(330,250){\special{em:lineto}}
\put(250,250){\special{em:moveto}}
\put(270,250){\special{em:lineto}}
\put(270,270){\special{em:lineto}}
\put(250,270){\special{em:lineto}}
\put(250,250){\special{em:lineto}}
\put(270,270){\special{em:lineto}}
\put(150,270){\special{em:moveto}}
\put(170,270){\special{em:lineto}}
\put(170,250){\special{em:lineto}}
\put(150,250){\special{em:lineto}}
\put(150,270){\special{em:lineto}}
\put(170,250){\special{em:lineto}}
\put(150,370){\special{em:moveto}}
\put(170,350){\special{em:lineto}}
\put(50,350){\special{em:moveto}}
\put(70,370){\special{em:lineto}}
\put(-10,260){\special{em:moveto}}
\put(-100,260){\special{em:lineto}}
\put(10,260){\special{em:moveto}}
\put(310,260){\special{em:lineto}}
\put(260,260){\special{em:moveto}}
\put(260,310){\special{em:lineto}}
\put(60,260){\special{em:moveto}}
\put(60,360){\special{em:lineto}}
\put(60,360){\special{em:moveto}}
\put(160,360){\special{em:lineto}}
\put(160,260){\special{em:moveto}}
\put(160,360){\special{em:lineto}}
\put(25,255){\special{em:moveto}}
\put(30,260){\special{em:lineto}}
\put(25,265){\special{em:lineto}}
\put(-25,255){\special{em:moveto}}
\put(-30,260){\special{em:lineto}}
\put(-25,265){\special{em:lineto}}
\put(85,255){\special{em:moveto}}
\put(90,260){\special{em:lineto}}
\put(85,265){\special{em:lineto}}
\put(155,310){\special{em:moveto}}
\put(160,305){\special{em:lineto}}
\put(165,310){\special{em:lineto}}
\put(55,285){\special{em:moveto}}
\put(60,290){\special{em:lineto}}
\put(65,285){\special{em:lineto}}
\put(-55,285){\special{em:moveto}}
\put(-60,290){\special{em:lineto}}
\put(-65,285){\special{em:lineto}}
\put(220,255){\special{em:moveto}}
\put(225,260){\special{em:lineto}}
\put(220,265){\special{em:lineto}}
\put(40,270){\special{em:moveto}}
\put(44,270){\special{em:lineto}}
\put(42,270){\special{em:moveto}}
\put(42,350){\special{em:lineto}}
\put(44,350){\special{em:moveto}}
\put(40,350){\special{em:lineto}}
\put(180,270){\special{em:moveto}}
\put(176,270){\special{em:lineto}}
\put(178,270){\special{em:moveto}}
\put(178,350){\special{em:lineto}}
\put(176,350){\special{em:moveto}}
\put(180,350){\special{em:lineto}}
\put(320,260){\makebox(0,0){\tiny$D$}}
\put(110,270){\makebox(0,0){\tiny$H$}}
\put(70,310){\makebox(0,0){\tiny$V$}}
\put(290,235){\makebox(0,0){\tiny$\phi = - {\pi \over 4}$}}
\put(20,300){\makebox(0,0){\tiny${\epsilon - \gamma \over 2}$}}
\put(200,300){\makebox(0,0){\tiny${\epsilon - \gamma \over 2}$}}
\put(70,238){\special{em:moveto}}
\put(70,242){\special{em:lineto}}
\put(70,240){\special{em:moveto}}
\put(150,240){\special{em:lineto}}
\put(150,238){\special{em:moveto}}
\put(150,242){\special{em:lineto}}
\put(110,230){\makebox(0,0){\tiny${\gamma}$}}
\put(0,260){\makebox(0,0){\tiny$S$}}
\put(10,250){\special{em:moveto}}
\put(-10,250){\special{em:lineto}}
\put(-10,270){\special{em:lineto}}
\put(10,270){\special{em:lineto}}
\put(10,250){\special{em:lineto}}
\put(-50,250){\special{em:moveto}}
\put(-70,250){\special{em:lineto}}
\put(-70,270){\special{em:lineto}}
\put(-50,270){\special{em:lineto}}
\put(-50,250){\special{em:lineto}}
\put(-70,270){\special{em:lineto}}
\put(-90,55){\special{em:moveto}}
\put(-110,65){\special{em:lineto}}
\put(-91,53){\special{em:moveto}}
\put(-111,63){\special{em:lineto}}
\put(-100,260){\special{em:moveto}}
\put(-220,140){\special{em:lineto}}
\put(-60,260){\special{em:moveto}}
\put(-60,310){\special{em:lineto}}
\put(-213,187){\special{em:moveto}}
\put(-227,173){\special{em:lineto}}
\put(-241,187){\special{em:lineto}}
\put(-227,201){\special{em:lineto}}
\put(-213,187){\special{em:lineto}}
\put(-250,197){\makebox(0,0){\tiny$D'$}}
\put(-150,214){\special{em:moveto}}
\put(-150,210){\special{em:lineto}}
\put(-146,210){\special{em:lineto}}
\put(-60,40){\makebox(0,0){\tiny$A$}}
\put(-60,240){\makebox(0,0){\tiny$A$}}
\put(-200,180){\makebox(0,0){\tiny$B$}}
\put(-90,70){\makebox(0,0){\tiny$H$}}
\put(-90,250){\makebox(0,0){\tiny$H$}}
\put(-40,90){\makebox(0,0){\tiny$V$}}
\put(-40,290){\makebox(0,0){\tiny$V$}}
\end{picture}

\caption{Two pairs of maximally entangled photons are created simultaneously in the two sources marked with S. Of each pair, one photon is sent through a polarizing beamsplitter (A) and then reunited with its counterpart from the other pair, in a conventional beamsplitter (B). A detector (D') is placed behind one of the exits of this beamsplitter. If it fires, the photons to the right of the sources are projected into the desired state. (Note that in an actual experiment the two pairs would most likely originate from one LBO crystal.) The other two photons each traverse a copy of the weak measurement setup described in Fig. \ref{wmfig}, with $\theta = 0$ and $\phi = - {\pi \over 4}$. The detectors D both click in the case of a successful bipartite postselection. For the single particle measurements only the weak measurement setup in the path of the photon to be measured is used, and the other is omitted.}\label{combfig}
\end{figure}
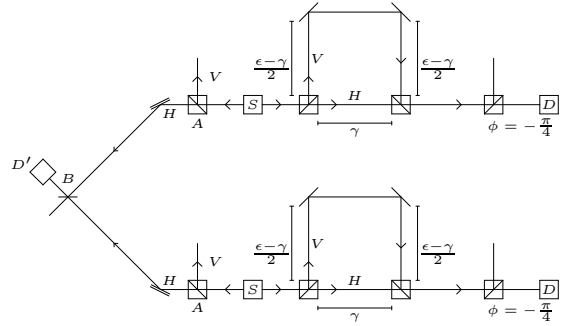

We can now apply our weak measurement setup to photons 2 and 4. The complete setup is shown in Fig. \ref{combfig}. 

For postselection we choose the same state as Aharonov {\em et al.} \cite{popescu}, namely the one corresponding to a simultaneous firing of the $D^\pm$ detectors in Hardy's original paradox:

\begin{eqnarray}\label{prior}
| \Psi_f \rangle  = {1 \over 2} [| H_2 H_4 \rangle - | H_2 V_4 \rangle - | V_2 H_4 \rangle + | V_2 V_4 \rangle]
\end{eqnarray} 

This is achieved by setting the basis angles of the two post-selecting polarizing beamsplitters to $\phi = -{\pi \over 4}$. 
The weak measurement operator for combined measurement of both photons is given \cite{us} by:

\begin{eqnarray}\label{weakop}
A_c^{2,4} = (\gamma, \gamma) |H_2H_4 \rangle \langle H_2H_4 | 
\cr
+ (\gamma, \epsilon) | H_2V_j \rangle \langle H_2V_j | 
\cr
+ (\epsilon, \gamma) | V_2H_j \rangle \langle V_2H_j | 
\cr
+ (\epsilon, \epsilon) | V_2V_j \rangle \langle V_2V_j | 
\end{eqnarray}

whereas the weak measurement operator for measurement of a single photon $2$ is given \cite{us} by:

\begin{eqnarray}
A_s^{2,4} = [\gamma | H_2 \rangle \langle H_2 | + \epsilon | V_2 \rangle \langle V_2 |] \otimes {\rm I}_{4}  
\end{eqnarray}

and similarly the operator for a single photon $4$ is:

\begin{eqnarray}
A_s^{2,4} = {\rm I}_{2} \otimes [\gamma | H_4 \rangle \langle H_4 | + \epsilon | V_4 \rangle \langle V_4 |] 
\end{eqnarray}

where $\gamma$ and $\epsilon$ are the delays in arrival time experienced by the photons for horizontal and vertical polarization respectively. (Note that $\gamma = 0$ in \cite{us}). ${\rm I}_2$ and ${\rm I}_4$ are the identity operators for the Hilbert spaces of photons $2$ and $4$ respectively. In the case of the combined measurement we have written the arrival times in form of a two-vector. 

Hence the weak values of the arrival times are as follows:

Only measuring photon 2: 

\begin{eqnarray}
A^2_w = {\langle \Psi_f | A_s^{2,4} | \Psi_i \rangle \over \langle \Psi_f | \Psi_i \rangle} = \epsilon 
\end{eqnarray}

Similarly, for photon 4 only:

\begin{eqnarray}
A^4_w = {\langle \Psi_f | A_s^{4,2} | \Psi_i \rangle \over \langle \Psi_f | \Psi_i \rangle} = \epsilon 
\end{eqnarray}

In turn, the combined measurement of photons 2 and 4 yields:

\begin{eqnarray}\label{combweak}
A^{2,4}_w = {\langle \Psi_f | A_c^{2,4} | \Psi_i \rangle \over \langle \Psi_f | \Psi_i \rangle} = (\gamma, \epsilon) + (\epsilon, \gamma) - (\gamma, \gamma) = (\epsilon, \epsilon)
\end{eqnarray}

These results correspond exactly to the weak values of the particle occupation numbers given by eqns. (\ref{pon}) and (\ref{pon2}) following Aharonov {\em et al.} in \cite{popescu}. However, in addition to these values,  eq. (\ref{combweak}) reveals in a very straightforward way how the three weak values $(\gamma, \epsilon)$, $(\epsilon, \gamma)$ and $- (\gamma, \gamma)$ - which correspond directly to the values in eq. (\ref{pon2}) - combine to give the paradoxical result $(\epsilon, \epsilon)$. It is paradoxical because $(\epsilon, \epsilon)$ is the result corresponding to the combined polarization state $|V_2 V_4 \rangle$, which was projected out. 

In other words, this setup shows how three unusual yet non-paradoxical pointer positions of the measurement apparatus form a superposition which corresponds to a paradoxical outcome of the measurement. 

In order to measure the three terms in eq. \ref{combweak} separately, one would have to turn the operator given in eq. \ref{weakop} into projectors such as $A^{2,4}_{c,HH} = (\gamma, \gamma) |H_2 H_4 \rangle \langle H_2 H_4 |$ or $A^{2,4}_{c,HV} = (\gamma, \epsilon) |H_2 V_4 \rangle \langle H_2 V_4 |$. This can be done by inserting two detectors, one between the first and second beamsplitters of each of the weak measurement setups - e.g. for $A^{2,4}_{c,HH}$ one would insert them in both $V$ channels. Thus if {\em neither} of these click, the state is projected according to the operator $A^{2,4}_{c,HH}$. 
\section{Conclusion}

We have proposed an experimental implementation of the analysis by Aharonov {\em et al.} \cite{popescu} of the gedankenexperiment known as Hardy's Paradox \cite{hardy}. In this setup the arrival times of photons act as pointer positions for the weak measurement of polarization states. The weak values predicted for the particle occupation numbers in \cite{popescu} - including the extraordinary negative ones - appear in this setup. Moreover our implementation reveals how these pointer positions combine to give rise to the paradoxical pointer position upon which Hardy's original paradox rests. 

S. E. Ahnert was supported by the Howard Research Studentship of Sidney Sussex College, Cambridge.

\end{multicols} 
\end{document}